\documentclass{article} \usepackage {latexsym, amsfonts, amsmath,
amsthm, graphics, amssymb, times, fullpage}

\makeatother

\newcommand{\set}[1]{\{#1\}}
\newcommand{\setcomprehension}[2]{\{#1 \mid{ } #2\}}
\newcommand{\nue}{\nu_{e}}

\newcommand{\numubar}{\overline{\nu}_{\mu}}
\renewcommand{\vec}[1]{\mathbf{#1}}
\newcommand{\q}{\mathbf{q}}
\renewcommand{\r}{\mathbf{r}}
\newcommand{\R}{\mathbf{R}}

\begin{document}
\title{Algorithmic Derivation of Additive Selection Rules and Particle Families from Reaction Data}
\author{Oliver Schulte and Mark S. Drew}
\date{\today}
\maketitle

\newtheorem{definition}{Definition} \newtheorem{example}{Example}
\newtheorem{theorem}{Theorem}
\newtheorem{lemma}[theorem]{Lemma} 
\newtheorem{corollary}[theorem]{Corollary}
\newtheorem{proposition}[theorem]{Proposition}
\newtheorem{finding}{Finding}

\begin{abstract}
  
  We describe a machine-learning system that uses linear
  vector-space based techniques for inference from observations to
  extend previous work on model construction for particle physics
  \cite{valdes-perez96:_system_gener_const_model_partic_famil,valdes-perez94:_system_induc_parsim_phenom_conser_laws,kocabas91:_confl_resol_discov_partic_physic}. The program searches for quantities conserved in all reactions from a given input set; given data based on frequent decays it rediscovers the family conservation laws: baryon\#, electron\#, muon\# and tau\#. We show that these families are uniquely determined by frequent decay data.
 \end{abstract}

\section{Introduction: Automated Search for Conserved Quantities}

One of the 
goals of particle physics theory is to find symmetries in particle interactions. This challenge has led to  the discovery  of new conservation laws for particle interactions. An important class of laws are {\em additive conservation laws} or {\em selection rules}. These \emph{selection
  rules}  are based on quantum numbers, physical quantities assigned to each particle. Table \ref{table:quant} shows the values of important quantum numbers for a set of particles \cite{PDBook}. The Standard Model with massless neutrinos includes the conservation of these quantum numbers  \cite{wolfenstein:_tests_conser_laws}. For brevity, in the following we refer to the quantities \textbf{\{charge, baryon, electron, muon, tau\}} denoted \textbf{\{C;BEMT\}} as the ``standard model quantities".

\begin{table}[tbp] \centering
\small
\begin{tabular}{|c|c|c|c|c|c|c|}
\hline
& Particle & Charge & Baryon\# & Tau\# & Electron\# & Muon\# \\ \hline
1 & $\Sigma ^{-}$ & -1 & 1 & 0 & 0 & 0 \\ \hline
2 & $\overline{\Sigma }^{+}$ & 1 & -1 & 0 & 0 & 0 \\ \hline
5 & $n$ & 0 & 1 & 0 & 0 & 0 \\ \hline
6 & $\overline{n}$ & 0 & -1 & 0 & 0 & 0 \\ \hline
7 & $p$ & 1 & 1 & 0 & 0 & 0 \\ \hline
8 & $\overline{p}$ & -1 & -1 & 0 & 0 & 0 \\ \hline
13 & $\pi ^{+}$ & 1 & 0 & 0 & 0 & 0 \\ \hline
14 & $\pi ^{-}$ & -1 & 0 & 0 & 0 & 0 \\ \hline
15 & $\pi ^{0}$ & 0 & 0 & 0 & 0 & 0 \\ \hline
16 & $\gamma $ & 0 & 0 & 0 & 0 & 0 \\ \hline
17 & $\tau ^{-}$ & -1 & 0 & 1 & 0 & 0 \\ \hline
18 & $\tau ^{+}$ & 1 & 0 & -1 & 0 & 0 \\ \hline
19 & $\nu _{\tau }$ & 0 & 0 & 1 & 0 & 0 \\ \hline
20 & $\overline{\nu }_{\tau }$ & 0 & 0 & -1 & 0 & 0 \\ \hline
21 & $\mu ^{-}$ & -1 & 0 & 0 & 0 & 1 \\ \hline
22 & $\mu ^{+}$ & 1 & 0 & 0 & 0 & -1 \\ \hline
23 & $\nu _{\mu }$ & 0 & 0 & 0 & 0 & 1 \\ \hline
24 & $\overline{\nu }_{\mu }$ & 0 & 0 & 0 & 0 & -1 \\ \hline
25 & $e^{-}$ & -1 & 0 & 0 & 1 & 0 \\ \hline
26 & $e^{+}$ & 1 & 0 & 0 & -1 & 0 \\ \hline
27 & $\nu _{e}$ & 0 & 0 & 0 & 1 & 0 \\ \hline
28 & $\overline{\nu }_{e}$ & 0 & 0 & 0 & -1 & 0 \\ \hline
\end{tabular}
\caption{Some Common Particles and Quantum Number
Assignments\label{table:quant}}
\end{table}

In this paper we present an algorithm for discovering conserved quantities from given particle reaction data provided by the user. Our methods are based on new techniques for machine learning in linear spaces, drawing on several new theorems in linear algebra. The goal of our system is to facilitate data exploration and automated model construction  ~\cite{alpaydin04:_introd_machin_learn_adapt_comput_machin_learn,langley87:_scien_discov,bib:raul-matrix}.

We apply our system to investigate selection rules for data based on frequent decay modes. These reactions do not include neutrino oscillations or chiral anomalies. Historically, the standard model selection rules were discovered incrementally by adding new rules in response to more evidence \cite{bib:ford}, \cite{bib:omnes}. Our program starts fresh and looks for a set of selection rules that is optimal for the input data. With the aid of the program, we can systematically explore alternative rules and investigate which features of the current conservation laws are particular and which are invariant.



We prove several mathematical theorems that apply to any class of reactions (e.g., strong interactions, weak interactions, all allowed interactions). In contrast, the findings of our data analysis hold only for reactions in our data set which is based on frequent decays. This paper describes the following results:

\begin{enumerate}
\item {\em For any class of reactions,} the number of independent quantities conserved
  in the reaction class is no greater than the number of particles with no
  decay mode in the class.
\item For the frequent decay reactions considered in our dataset, the standard model quantities \textbf{\{C;BEMT\}} are complete in the sense that every other quantity conserved in these reactions is a linear combination of \textbf{\{C;BEMT\}}.

\item The particle families corresponding to \textbf{\{BEMT\}} are uniquely determined by the frequent decay data in our data set.

\end{enumerate}

The next section presents our algorithm and findings, followed by a summary. We briefly discuss applying our data analysis system to reactions that involve neutrino oscillation and/or chiral anomalies. Section \ref{sec:methods} describes our dataset and gives formal proofs of the new linear algebra theorems in this paper.

\section{Algorithm and Results} \label{sec:results}

We represent reactions and quantum numbers as $n$-vectors
\cite{valdes-perez94:_system_induc_parsim_phenom_conser_laws} 
\cite{aris69:_elemen_chemic_react_analy}, with the known particles numbered as $p_1,\dots, p_n$. Given a
reaction $r$, we subtract the sum of occurrences of $p_i$ among the
products from the sum among the reagents to obtain the \emph{net
  occurrence} of particle $p_i$ in reaction $r$. For example, in the
transition $p + p \rightarrow p + p + \pi^0$, the net occurrence of $p$
is $0$ and that of $\pi$ is  $-1$. The $n$-vector $\vec{r}$ is then defined
by setting $\vec{r}(i) =$ the net occurrence of $i$, where $\r(i)$ denotes the $i$-th component of the $n$-vector $\r$. Since net occurrences
are integers, we refer to $n$-vectors with integer entries as
\emph{reaction vectors}.  

A quantum number is represented by an $n$-vector $\vec{q}$, where $\vec{q}(i) =$
the quantum number for particle $p_i$. For example, if particle $p_1$ 
is $\Sigma^-$, then \textbf{charge}$(1) =-1$. If $\vec{q}$ is a quantum
number and $\vec{r}$ a reaction, then $\vec{q}$ is conserved in
$\vec{r}$ iff $\vec{q}\cdot \vec{r} = 0$. If we write $\mathbf{E}$ for the set of
input vectors that represent experimentally established reactions, \emph{the space
of quantities conserved in all reactions in $\mathbf{E}$ is the
orthogonal complement $\mathbf{E}^\bot$}.  We generally write $E$ for a set of input reactions to be analyzed by our algorithm, and $R$ for an arbitrary class of reactions for which we prove a mathematical theorem. Table \ref{table:encode} illustrates these concepts.

\begin{table}[tbp] \centering
\begin{tabular}
[c]{|l|l|l|l|l|l|l|l|}\hline%
\begin{tabular}
[c]{l}%
Particle\\\hline
Process/Quantum Number
\end{tabular}
&
\begin{tabular}
[c]{l}%
1\\
$p$%
\end{tabular}
&
\begin{tabular}
[c]{l}%
2\\
$\pi^{0}$%
\end{tabular}
&
\begin{tabular}
[c]{l}%
3\\
$\mu^{-}$%
\end{tabular}
&
\begin{tabular}
[c]{l}%
4\\
$e^{+}$%
\end{tabular}
&
\begin{tabular}
[c]{l}%
5\\
$e^{-}$%
\end{tabular}
&
\begin{tabular}
[c]{l}%
6\\
$v_{\mu}$%
\end{tabular}
&
\begin{tabular}
[c]{l}%
7\\
$\overline{\nu}_{e}$%
\end{tabular}
\\\hline
$\mu^{-}\rightarrow e^{-}+\nu_{\mu}+\overline{\nu}_{e}$ &
\multicolumn{1}{|c|}{0} & \multicolumn{1}{|c|}{0} & \multicolumn{1}{|c|}{1} &
\multicolumn{1}{|c|}{0} & \multicolumn{1}{|c|}{-1} & \multicolumn{1}{|c|}{-1}
& \multicolumn{1}{|c|}{-1}\\\hline
$p\rightarrow e^{+}+\pi^{0}$ & \multicolumn{1}{|c|}{1} &
\multicolumn{1}{|c|}{-1} & \multicolumn{1}{|c|}{0} & \multicolumn{1}{|c|}{-1}
& \multicolumn{1}{|c|}{0} & \multicolumn{1}{|c|}{0} & \multicolumn{1}{|c|}{0}%
\\\hline
$p+p\rightarrow p+p+\pi^{0}$ & \multicolumn{1}{|c|}{0} &
\multicolumn{1}{|c|}{-1} & \multicolumn{1}{|c|}{0} & \multicolumn{1}{|c|}{0} &
\multicolumn{1}{|c|}{0} & \multicolumn{1}{|c|}{0} & \multicolumn{1}{|c|}{0}%
\\\hline
Baryon Number & \multicolumn{1}{|c|}{1} & \multicolumn{1}{|c|}{0} &
\multicolumn{1}{|c|}{0} & \multicolumn{1}{|c|}{0} & \multicolumn{1}{|c|}{0} &
\multicolumn{1}{|c|}{0} & \multicolumn{1}{|c|}{0}\\\hline
Electric Charge & \multicolumn{1}{|c|}{1} & \multicolumn{1}{|c|}{0} &
\multicolumn{1}{|c|}{-1} & \multicolumn{1}{|c|}{1} & \multicolumn{1}{|c|}{-1}
& \multicolumn{1}{|c|}{0} & \multicolumn{1}{|c|}{0}\\\hline
\end{tabular}
\caption{The Representation of Reactions and Quantum Numbers as $n$-vectors}\label{table:encode}%
\end{table}%

A fundamental principle that has guided the search for conservation
principles in particle physics was dubbed Gell-Mann's Totalitarian
Principle: 
``Anything which is not
prohibited is
compulsory''\cite{bilaniuk69:_partic_beyon_light_barrier}. Thus if a
reaction is not observed, some physical law must forbid it. The more
linearly independent quantum numbers we introduce, the more unobserved
processes our conservation principles rule out. Hence we seek a \emph{basis
in the nullspace of the observed reactions}.  
%
Our analysis considers
a comprehensive set of $n=193$
particles (see Section \ref{sec:design-part-react} for details). Reactions that are linear combination of previous observed reactions (when viewed as $n$-vectors) do not lead to new constraints on additive selection rules. So we seek a
maximal set of linearly independent
observed reactions, which the next
proposition helps us find.  

\begin{proposition}
  \label{pr:linear-indep}
  Let $\vec{d_1},\dots,\vec{d_k}$ be a set of vectors for decays of
  distinct particles. That is, $\vec{d_i}$ is a process of the form
  $p_i\rightarrow\dots$. Then conservation of energy and momentum
  imply that the set $\set{\vec{d_1},\dots, \vec{d_k}}$ is linearly
  independent.   
\end{proposition}

 To illustrate the proposition, we note that for our 193 particles, there are established
 decay modes for all but $11$ particles (photon, proton, electron, the
 three neutrinos, and the respective antiparticles). The proposition
 guarantees that we obtain at least $k = 193-11 = 182$ linearly
 independent reaction vectors from these decay modes.

Since $dim(\R) + dim(\R^{\bot}) = n$, it follows that if $k$ is the number of particles with decay modes in a class of reactions $R$, then
$dim(\mathbf{R})\geq k$, so $dim(\mathbf{R}^\bot)\leq n-k$, which is
the number of particles without a decay mode in the reaction class $R$. Thus we have a
surprising relationship: 
\begin{corollary}
For any class of reactions $R$, the number of independent quantum numbers conserved in the class is bounded by the number of particles without a decay mode in the class. 
\end{corollary}

Note that this result holds for partial symmetries too  since it holds for any subclass of processes to which the symmetry applies.
To illustrate, since there are $11$ stable particles without any known decay modes, we know a priori
that there can be at most $11$ linearly independent quantities conserved
in all processes. In fact, we find a stronger necessary
relationship between the number of particles without a decay mode and the number of independent conserved quantities, if we take into account
the matter/antimatter division of elementary particles. 
Say that a
quantum number $\vec{q}$  is \emph{coupled} with respect to
matter/antimatter, or simply coupled, if $\vec{q}(i) = -\vec{q}(j)$ when particle
$p_i$ is the antiparticle of $p_j$. The quantum numbers \textbf{\{C;BEMT\}} are all coupled with respect to matter/antimatter (see Table \ref{table:quant}). The following proposition assumes
that particle $p$ has a decay mode just in case its antiparticle
$\overline{p}$ does. 

\begin{proposition} \label{prop:pairs}
Let $s$ be the number of particle/antiparticle pairs $(p, \overline{p})$ that have no decay mode in a class of reactions $R$. Then conservation of energy and momentum imply that there are at most $s$ coupled independent quantum numbers conserved in all reactions in the class.
\end{proposition}

There are $6$ particle/antiparticle pairs without a decay mode: $(p,
\overline{p})$, $(e^-, e^+)$, $(\gamma,\gamma)$, $(\nu_e,\overline{\nu_e})$,
$(\nu_\mu, \overline{\nu}_\mu)$, $(\nu_\tau, \overline{\nu}_\tau)$. Thus
the proposition implies that there can be at most $6$ coupled linearly
independent quantum numbers conserved in all allowed reactions.

Based on Proposition~\ref{pr:linear-indep}, we included in $\vec{E}$ one decay mode for
each particle that has one listed in the particle data Review
\cite{PDBook}, and a number (more than $11$) of other known reactions
resulting in a total of $205$ datapoints. On this data, our computation
establishes the following.  
\begin{finding}
The quantum numbers \textbf{\{C;BEMT\}} form a basis for the nullspace
$\mathbf{E}^\bot$ of the reaction dataset $\textbf{E}$ based on frequent decays. Therefore, any other
quantity conserved in all of these reactions is a linear combination of
\textbf{\{C;BEMT\}}.
\end{finding}

There are many sets of conserved quantities predictively equivalent to the standard model quantities \textbf{\{C;BEMT\}}. By ``predictive equivalence" we mean that any reaction $\r$ conserves all the quantities \textbf{\{C;BEMT\}} if and only if $\r$ conserves all alternative quantities. In vector terms, a set of alternative quantities $\{\q_1,...,\q_5\}$ is equivalent to \textbf{\{C;BEMT\}} if and only if both sets span the same linear space. Although many alternative theories forbid the same reactions, the basis \textbf{\{C;BEMT\}} has a
special feature that singles it out. The key insight is that these
quantities not only classify reactions into
``forbidden'' and ``allowed'',
but also group particles into families. Say that particle $p_i$ \emph{carries} a
quantum number $\vec{q}$ if $\vec{q}(i)\not=0$. For example, the
carriers of electron number are the electron, positron, electron
neutrino, electron antineutrino (see Table \ref{table:quant}). A set of quantum numbers
$\set{\vec{q_1},\dots,\vec{q_m}}$ forms {\em a family set} if no particle carries two
quantities; formally, if $\vec{q_j}(k) = 0$ whenever $\vec{q}_i(k)\not=0$, for
all $i\not=j$. The quantum numbers \textbf{BEMT} form a family set. 

\begin{theorem}
  \label{thm:famil-set}
Let $\{\vec{q_1, q_2, q_3, q_4}\}$ be \emph{any family set} of quantum
numbers such that $\{\vec{q_1, q_2, q_3, q_4, C}\}$ is predictively equivalent to the standard model quantities \textbf{\{C;BEMT\}}. Then for each $\vec{q_i}, i=1..4$, there is a standard quantity from
\textbf{\{BEMT\}} such that the carriers of $\vec{q}_i$ are the carriers of the standard
quantity. In other words, any such family set $\vec{q_1,q_2,q_3,q_4}$ determines
the same particle families as \textbf{\{BEMT\}}.   
\end{theorem}

This result can be interpreted as showing that the particle families corresponding to the baryons and the three lepton generations are {\em invariant} with respect to different alternative assignments of quantum numbers: any assignment of quantum numbers that is (1) predictively equivalent to the quantities \textbf{\{C;BEMT\}}, and (2) based on a division of particles into any families must in fact be based on the baryon family and the three lepton generations. Section \ref{sec:proof-that-particle} shows that the uniqueness of particle families is a general fact that holds not just for the families of the Standard Model.
 
We applied Theorem \ref{thm:famil-set} to computationally rediscover
the \textbf{\{BEMT\}} quantum number assignments: a computational
search for an extension $\mathbf{Q}$ of $\{\mathbf{C}\}$ to a basis for
$\mathbf{E}^\bot$ that  minimizes the sum of the absolute
values of the quantum numbers yields \textbf{\{BEMT\}} as the
solution (up to sign). Theorem~\ref{thm:famil-set} establishes a tight relationship
between particle dynamics and particle taxonomy: a given particle
taxonomy suggests an explanation of reaction data via family
conservation laws such as \textbf{\{BEMT\}}; conversely the reaction data
can be used to \emph{find} a unique taxonomy corresponding to a complete
set of conserved quantities. We emphasize that the program rediscovers the baryon family and the three generations of leptons from reaction data alone, without any knowledge of particle families or particle properties at all; internally, the program represents a particle simply as a natural number.

\section{Summary and Further Applications}
We described a new algorithm for finding an optimal set of selection rules for given reaction data and several linear algebra theorems that provide analytic insight into properties of selection rules. We applied the algorithm to a set of observed transitions consisting mainly of frequent decay modes. Our computations in combination with the mathematical analysis yield the following results. (1) For any class of reactions, the number of irredundant selection rules is bounded above by the number of particles without a decay mode in the class (counting particle-antiparticle pairs just once). (2) The 
quantities \textbf{\{C;BEMT\}} (= charge, baryon number, electron number, muon number, tau number) are optimal for our data set based on frequent decays in that they explain the nonoccurrence of as many unobserved processes as possible.
(3) Given the conservation of electric charge, any optimal set of selection rules  for our data set that is based on dividing particles into families must correspond to the particle families defined by the baryon, electron, muon and tau quantum numbers. Thus these families are uniquely determined by the data. 

Since our data set includes the most probable decay mode for each particle that has one, it excludes low-frequency events such as neutrino oscillations and chiral anomalies. As our algorithms can be  used to find symmetries in any class of interactions, there is no obstacle in principle to apply them to data sets that include these types of events. For example, we could extend our data set $E$ to include rare decays such as $\mu \rightarrow e^- + \nue+ \numubar$. On this input data we expect the algorithm to find the conservation of electric charge, baryon number, and lepton number. Additional input processes could include reactions that violate the conservation of baryon number; in that case our hypothesis is that the algorithm will indicate the conservation of electric charge and baryon-lepton number.

\section{Methods and Proofs}
\label{sec:methods}

\subsection{Design of the Particle and Reaction Database}
\label{sec:design-part-react}

 The $193$-particle database is a comprehensive catalog of the known
 particles; some particles were excluded by the following
 criteria. (1) The database contains only particles included in the
 summary table \cite{PDBook}, which excludes some particles whose
 existence and properties needs further confirmation. (2) We omitted
 some resonances. (3) We did not include quarks. (4) We included separate entries for each particle and its antiparticle, for example the proton $p$ and its antiparticle $\overline{p}$ are listed separately. The reason is to see if the program can rediscover from the reaction data which particle pairs behave like antiparticles of each other, which it does. The complete particle and reaction databases are available in Excel format at http://www.cs.sfu.ca/~oschulte/particles/ , and a list of all included particles at http://www.cs.sfu.ca/~oschulte/particles/particle-list.txt .

Proposition~\ref{pr:linear-indep} is the rationale for basing our reaction
database on decays. The proof is as follows. Without loss of generality, assume
that particles are numbered by mass in descending order, so that
$\mbox{mass}(p_i) \geq \mbox{mass}(p_j)$ whenever $i\leq j$. It is well-known
that conservation of energy and momentum implies that if $\vec{d}_i = p_i
\rightarrow p^0 + p^1 + p^2 + \cdots+ p^m$ is a possible decay, then
$\mbox{mass}(p^j) <\mbox{mass}(p_i)$ for $j=0,..,m$. Let $D$ be the matrix whose rows are the vectors
$\vec{d}_1,\dots, \vec{d}_k$ representing the decays of distinct particles. Fix
$i$ and consider $j < i$; then $\mbox{mass}(p_j)\geq \mbox{mass}(p_i)$, and so
particle $p_j$ does not occur in decay $d_i$; hence $D_{ij} = 0$. Since this
holds for arbitrary $i$, $j<i$, it follows that $D$ is upper triangular and
so the set of its row vectors is linearly independent. 

We omit the proof of Proposition \ref{prop:pairs} which has the same basic idea. 

\subsection{Determination of Particle Families by Reaction Data}
\label{sec:proof-that-particle}

We first show that any two family bases have the same carriers. Recall
that $\mbox{carriers}(\vec{v})=\{i | \vec{v}(i)\neq 0\}$, and
that $B$ is a family basis if for $\vec{v}_1, \vec{v}_2 \in B$ we have
carriers$(\vec{v}_1)\cap$carriers$(\vec{v}_2)=\emptyset$. Say that a
basis $B'$ is a \emph{multiple} of a basis $B$ if for every vector $\vec{v}'$ in $B'$,
there is a vector $\vec{v}$ in $B$ and a scalar $a$ such that $\vec{v}'= a\vec{v}$. If $\vec{v}
= a\vec{v}'$ for $a\not=0$, then $\vec{v}$ and $\vec{v}'$ have the
same carriers, so the bases $B$ and $B'$ determine the same families if
$B'$ is a multiple of a family basis $B$.  

\begin{proposition}
Suppose that $B, B'$ are family bases for a linear space $V$. Then $B'$ is a multiple of $B$.
\end{proposition}
\begin{proof}
Let $\vec{v}'$ be a vector in $B'$. Then we may write $\vec{v}' =
\sum^n_{i=1} a_i \vec{v}_i$, where $\vec{v}_i \in B$, with
$n= dim(B)$. Since $B'$ is a basis, $\vec{v}'\not=\vec{0}$ and
there exists $a_i\not= 0$. Then $$\mbox{carriers}(\vec{v}_i)\subseteq \mbox{carriers}(\vec{v}')$$
because $B$ is a family set. We may write $$\vec{v}_i =\sum^n_{k=1}
b_k \vec{v}'_k + b \vec{v}',$$ where $\vec{v}'_k\not= \vec{v}'$ is in
$B'$. Since $B'$ is a family set, the carriers of $\sum^n_{k=1} b_k
\vec{v}'_k$ are disjoint from those of $\vec{v}'$. As the
carriers of $\vec{v}'$ include those of $\vec{v}_i$, it follows that $\vec{v}_i = b
\vec{v}'$ where $b\not= 0$, and so $\vec{v}' = \frac{1}{b} \vec{v}_i$. Since this
holds for any vector in $B'$, it follows that $B'$ is a
multiple of $B$. 
\end{proof}

The basis \textbf{\{C;BEMT\}} is not a family basis for
$\mathbf{E}^\bot$
because the carriers of electric charge
$\mathbf{C}$ occur among the carriers of all other conserved
quantities. Electric charge has the special status of being
\emph{logically independent} of the other quantities \textbf{\{BEMT\}};
we define this notion as follows. A quantity $\vec{q}$ is
logically independent of a set of quantities $B$ if for all $\vec{v}$ in $B$
we have that (1) some carrier of $\vec{v}$ is not a carrier of $\vec{q}$, (2)
some carrier of $\vec{q}$ is not a carrier of $\vec{v}$, and (3) some
particle carries both $\vec{p}$ and $\vec{q}$. 

\begin{theorem} \label{theo:general-family}
Let $B\cup\vec{q}$ be a basis for $V$ such that $|B| > 2$, $B$ is a
family set and $\vec{q}$ is logically independent of $B$. Let $B'$
be another family set such that $B'\cup\set{\vec{q}}$ is a basis for
$V$. Then $B'$ is a linear multiple of $B$. 
\end{theorem}

\begin{proof}
Let $\vec{v}_q$ be the vector $\vec{v}$ with $0$ assigned to all
carriers of $\vec{q}$ (i.e. $\vec{v}_q(i) = 0$ if $\vec{q}(i) = 0$,
and $\vec{v}_q(i) = \vec{v}(i)$ otherwise). For a set of vectors $U$,
let $U_q = \setcomprehension{\vec{v}_q}{\vec{v}\in U}$. It is easy to
verify that $B_q$ and $B'_q$ are bases for $V_q$, so the previous
Proposition implies that $B'_q$ is a multiple of $B_q$. Thus every
quantity $\vec{v}'_i$ in $B'$ is of the form $$\vec{v}'_i = a_i\vec{v}_i + b_i\vec{q}$$ for some
vector $\vec{v}_i \in B$; we argue by contradiction that $b_i = 0$ for all
$\vec{v}_i' \in B'$, which establishes the theorem.

Case 1: there are two distinct $\vec{v}_i, \vec{v}_j$ such that $b_i
\not= 0$ and $b_j \not=0$. Since $|B| > 2$, there is a $\vec{v}_k$
different from $\vec{v}_i, \vec{v}_j$. As $B$ is a family set and
$\vec{q}$ is logically independent of $B$, this implies that there is a
particle $p$ carrying both $\vec{v}_k$ and $\vec{q}$ but neither $\vec{v}_i$ nor
$\vec{v}_j$. So $p$ carries $\vec{v}'_i = a_i \vec{v}_i + b_i \vec{q}$ and
also $\vec{v}_j = a_j \vec{v}_j + b_j \vec{q}$, which contradicts the
supposition that $B'$ is a family set. 

Case 2: there is exactly one $\vec{v}'_i$ such that $b_i\not=
0$. Choose a vector $\vec{v}_j$ and a particle $p$ such that $p$
carries both $\vec{v}_j$ and $\vec{q}$, but not $\vec{v}_i$. Then $p$
carries $\vec{v}_i = a_i \vec{v}_i + b_i \vec{q}$, and since $\vec{v}'_j = a_j
\vec{v}_j$, the particle $p$ carries $\vec{v}'_j$ as well, which
contradicts the supposition that $B'$ is a family set. 

In either case we arrive at a contradiction, which shows that $B'$ is a linear multiple of $B$.
\end{proof}

Theorem \ref{thm:famil-set} follows immediately by setting $\vec{q} = {\mathbf C}$ and $B =$ {\bf BEMT}.

\section*{Acknowledgements}
\label{sec:acknowledgements}
We are grateful to Robert Coleman, Manuela Vincter, Matthew Strassler, Raul Vald{\'e}s-P{\'e}rez, Kevin Kelly and Clark Glymour for discussion. We presented these results to the Particle Physics Group at Simon Fraser University and received helpful comments. Alexandre Korolev set up the particle database and Gregory Dostatni wrote the first program version. This research was supported by grants to each author from the Natural Sciences and Engineering Research Council of Canada.

\bibliographystyle{plain}
\small
\bibliography{QuantumPhysics,oliver}
\end{document}